\documentclass{optica-article}
\usepackage{textcomp}

\journal{opticajournal} % for journals or Optica Open

\articletype{Research Article}

\usepackage{lineno}
\begin{document}

\title{Induced Coherence in Quantum and Classical Interferometry: Origin and Control}

\author{Balakrishnan Viswanathan \authormark{1,2 *}, Ravi Kamal Pandey \authormark{3}, Leena Barshilia \authormark{4}, Devendra Kumar Mishra \authormark{3}, Sibasish Ghosh \authormark{1,2} and Prasanta K. Panigrahi \authormark{4,5}}

\address{\authormark{1} Optics and Quantum Information Group, The Institute of Mathematical Sciences, C.I.T. Campus, Taramani, Chennai 600113, India\\
\authormark{2} Homi Bhabha National Institute, Training School Complex, Anushakti Nagar, Mumbai 400094, India \\
\authormark{3} Department of Physics, Institute of Science, Banaras Hindu University, Varanasi 221005, India\\
\authormark{4}Centre for Quantum Science and Technology, Siksha `O' Anusandhan, Bhubaneswar 751030, Odisha, India \\
\authormark{5} Department of Physical Sciences, Indian Institute of Science Education and Research
(IISER), Kolkata, Mohanpur 741246, India}

\email{\authormark{*}bala.krishvishwa@gmail.com} %% email address is required; see note below about the corresponding author designation

% use {asbstract*} to suppress the copyright line. Copyright information will be added in production

\begin{abstract*} 
The wave-particle duality that manifests in Young’s double-slit interferometry satisfying the well known Englert-Greenberger-Yasin inequality has been shown to obey a triality relation involving coherence, predictability and rather surprisingly entanglement.  In the present work, we explicitly demonstrate the persistence of the triality relation in the quantum induced coherence interferometer and the classical configuration recovers the well known duality inequality.  Furthermore, we explore the origin of coherence and entanglement, in the induced coherence interferometer, and means to control them.  We show that path indistinguishability (or intrinsic indistinguishability of the sources) in our setup plays a complementary role in influencing both interference and entanglement. We bring out manifestly the nature of coherence in both classical and quantum versions of induced coherence interferometers. 

\end{abstract*}

%%%%%%%%%%%%%%%%%%%%%%%%%%  body  %%%%%%%%%%%%%%%%%%%%%%%%%%
\section{Introduction}

Photons and electrons manifest wave and particle nature in their path degrees of freedom. It is a strange enigma that many experiments performed since the birth of quantum physics have revealed either the wave or particle nature but never both these aspects simultaneously. Bohr postulated this puzzling fact as a fundamental principle which is today known as the ``Complementarity Principle" \cite{Bohr1928TheQP, bohr1958atomic} whose twin ingredients are \textit{mutual exclusiveness} and \textit{joint completion}. ``Mutual exclusiveness'' means that we observe either the wave nature or the particle nature but never both of them concurrently and ``Joint completion'' says that these two attributes together give the complete description of a quantum system. A detailed discussion on Bohr's complementarity principle can be found in \cite{starke2026bohr}. Over time, the complementarity principle has been given a more quantitative structure starting with the influential work of Wootters and Zurek \cite{wootters1979complementarity} followed by several others \cite{glauber1986amplifiers,greenberger1988simultaneous, mandel1991coherence,jaeger1993complementarity,jaeger1995two, englert1996fringe,Liu2012_WaveParticle} the result of which is the following inequality for the duality relation 
\begin{align} \label{Duality-Eq}
 \mathcal{V}^{2} + \mathcal{P}^{2} &\leq 1,   
\end{align}
where $\mathcal{V}$ is the visibility of the interference fringes (corresponding to the wave nature of the quantum object of interest) and $\mathcal{P}$ is the predictability that gives the location of the quantum object (corresponding to the particle nature). 
\par
Since then, it has been shown that the duality inequality can be converted into a tight equality (that is a three-way identity relation) involving visibility, predictability and concurrence ($\mathcal{C}$), known as the triality relation \cite{jakob2010quantitative, Jakob2007Complementarity, qian2018entanglement, qian2020turning, roy2022coherence}:
\begin{align} \label{Triality-Eq}
 \mathcal{V}^{2} + \mathcal{P}^{2} + \mathcal{C}^{2} &= 1,   
\end{align}
which demonstrates that the wave-particle complementarity is intrinsically connected to entanglement between the path and other degrees of freedom (such as detector) of the same quantum system. The triality relation has been shown to persist in a variety of systems such as Young's double-slit setup with genuine point sources (pair of two-level atom) \cite{qian2020quantum}, open quantum systems \cite{swain2026persistence} and neutrino oscillations \cite{benerjee2026wave}.  
\par
One of the cardinal principles of quantum physics is that if there are two or more alternative ways for an event to happen with no means to distinguish between those alternatives, then one observes interference. Any successful attempt to gain knowledge of the alternatives destroys the interference. This profound yet perplexing principle, in the context of the double-slit experiment, was described by Feynman as a phenomenon ``\textit{...which has in it the heart of quantum mechanics. In reality, it contains the only mystery.}"\cite{feynman1966feynman}
\par
In the present work, we explore induced coherence interferometers, which operate on the principle of absence of ``which-source information" \cite{zou1991induced}, to investigate the origin of coherence and entanglement. We demonstrate the persistence of the triality relation in the quantum induced coherence interferometer while the classical induced coherence setup recovers the familiar duality relation. In this interferometric configuration, a coherent  superposition of the photon-pair-generation alternatives is created: a superposition of the origin of the photon pair itself. This setup involves sources producing two identically correlated pairs of photons. By overlapping the paths of one of the photons in the two pairs by the process of alignment, the information about the origin of the second photon is erased. This induces optical coherence in the second photon which then exhibits interference due to which this setup is called ``induced coherence interferometer''. The quality of interference of one of the photons in the pair depends on the extent to which the information about its origin is erased. The effect of interference vanishes completely when there is complete knowledge of the source of the photon (see \cite{hochrainer2022quantum} for a comprehensive review and \cite{lahiri2020partially} for a pedagogical exposition of induced coherence interferometry). 
\par
We quantitatively establish, in the induced coherence interferometer [see Fig.~(\ref{fig:ZWM-setup})], the relationship between path indistinguishability (or intrinsic degree of indistinguishability of the sources) and normalized mutual coherence, and how these two quantities are influenced by alignment. We look into the effect of path indistinguishability on interference and entanglement. In this context, the entanglement that we study is not in the internal degrees of freedom of photons such as polarization, spatial variables, orbital angular momentum etc. We are rather interested in mode entanglement, i.e., entanglement between the two path alternatives of the interfering photons [characterized by their respective modes: $S_{1}$ and $S_{2}$ in Fig.~(\ref{fig:ZWM-setup})] and the distinguishable modes of the companion photons in the pair [$I_{1}$ and $I_{2}$ in Fig.~(\ref{fig:ZWM-setup})] that are aligned (to achieve path indistinguishability). We show that the degree of indistinguishability influences both interference and entanglement in a complementary way viz., there is an inevitable trade-off between these two quantities in the induced coherence interferometer. 
\par
The origin of interference and mode entanglement, in our setup, is the coherent superposition of the two pair-generation alternatives. The coherence in the total quantum state manifests either as first-order coherence (interference) of one of the photons in the pair or as entanglement between the modes of  twin photons. Through our analysis in this work, we show that interference and mode entanglement are different manifestations of the same underlying quantum coherence. On the other hand, if the twin-photon pairs produced by the two sources end up in an incoherent mixture of states, we will neither observe interference nor will there be mode entanglement between the photons in the pair regardless of the degree of path indistinguishability. 
\par
Furthermore, we clearly spell out the role of alignment in both classical and quantum induced coherence interferometers, and bring out the similarities and fundamental differences on alignment between the two configurations. To sum up, we accomplish the following in this work:
\begin{itemize}
    \item We thoroughly investigate the quantum induced coherence interferometer, for both pure and mixed states, in the single-mode framework and expound the complementary role that the degree of path indistinguishability plays in influencing interference and mode entanglement. Along the way, we also verify the triality relation in this configuration. 
    \item We study the effect of an arbitrary background field (irrespective of its specific nature) on predictability, visibility and concurrence. 
    \item We clearly stress the role of alignment in both classical and quantum induced coherence interferometers, and bring out the subtle differences in both versions. 
\end{itemize}
\par
The paper is organized as follows: In Sec.~(\ref{sec:Q-ICI}), we discuss the quantum induced coherence interferometer in detail followed by a conceptual discussion on the role of alignment in classical and quantum induced coherence setups in Sec.~(\ref{sec:Classical ZWM}). We finally summarize our work in Sec.~(\ref{sec:concl}). 

\section{Quantum Induced Coherence Interferometer}
\label{sec:Q-ICI}

In this section, we first discuss the general framework of quantum induced coherence interferometer in Sec.~(\ref{subsec:Q-ICI-General}) where the underlying physical principles that operate this interferometer are discussed along with a detailed mathematical analysis. The notion of mode entanglement and the tight triality relation are treated in Sec.~(\ref{subsec: Q-ICI-Triality}). Following this, we explicitly show the equivalence between the traditional analysis of the induced coherence interferometer and the Hilbert-Schmidt coherence formalism in Sec.~(\ref{subsec: Q-ICI-HS Formalism}). In Sec.~(\ref{subsec: Q-ICI-Path Indistinguishability}), we quantitatively relate degree of path indistinguishability to mutual coherence (first-order correlation) of the interfering photons, and how path indistinguishability influences interference and mode entanglement in our interferometric setup. The effect of an arbitrary background field and its impact on visibility, predictability and concurrence are discussed in Sec.~(\ref{subsec: Q-ICI-Background Field}). The mixed state induced coherence interferometer is considered in Sec.~(\ref{subsec: Q-ICI-Mixed State}). 
\par
A different formalism to study complementarity in the induced coherence interferometer has been discussed in \cite{machado2024complementarity} in which distinguishability (regarding the source where the photon pair originated) is related to the second-order correlations between the twin photons. However, in our work considered here, we don't use second-order correlations at all. Our investigation is restricted to first-order correlation and the method of analysis that we have adopted in this work is very different from \cite{machado2024complementarity}. 

\subsection{General Framework}
\label{subsec:Q-ICI-General}

\begin{figure}
		\centering 
		\includegraphics[width=1.0\linewidth]{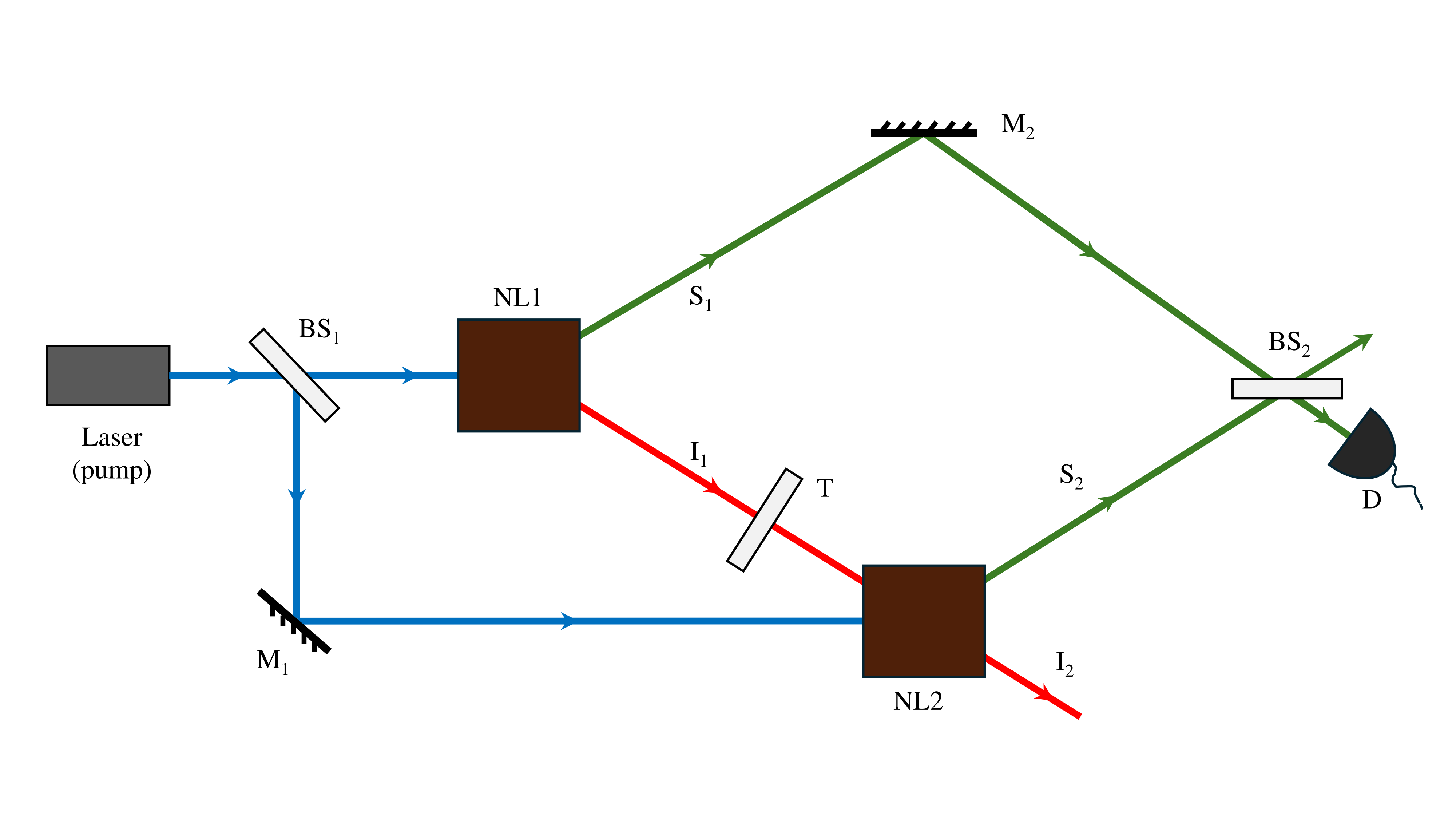}
		\caption{Schematic of quantum induced coherence interferometer: A weak laser passes through a $50-50$ beam splitter ($BS_{1}$), and pumps two identical nonlinear crystals $NL1$ and $NL2$. Each crystal, through spontaneous parametric down-conversion (SPDC), produces a pair of correlated photons which we call signal $S$ (green lines) and idler $I$ (red lines); $NL1$ and $NL2$ produce pairs $(S_{1},I_{1})$ and $(S_{2},I_{2})$, respectively. The idler $I_{1}$ from $NL1$ is aligned with the idler $I_{2}$ from $NL2$. The extent of alignment is controlled by a filter with variable transmission coefficient ($T$) placed between the two crystals. The two signal photons $S_{1}$ and $S_{2}$ pass through a $50-50$ beamsplitter $(BS_{2})$ and the interference pattern is recorded at the detector $D$. The idler is not detected in this scheme.}
		\label{fig:ZWM-setup}
	\end{figure}

We present a detailed analysis of the induced coherence interferometer with two identical pairs of correlated photons. We carry out our investigation in the single-mode framework for the pure state. The mixed state case is discussed in Sec.~(\ref{subsec: Q-ICI-Mixed State}). The schematic of the interferometric configuration is shown in Fig.~(\ref{fig:ZWM-setup}).
In this setup, two identical nonlinear crystals $NL1$ and $NL2$ are weakly pumped by a laser. The pump in Fig.~(\ref{fig:ZWM-setup}) is coherent. Each crystal, through spontaneous parametric down-conversion (SPDC), produces a pair of correlated photons which we label as signal ($S$) and idler ($I$); $NL1$ and $NL2$ produce pairs $(S_{1},I_{1})$ and $(S_{2},I_{2})$, respectively. Since the crystals are weakly pumped, the probability for both of them to emit simultaneously is negligible. The total quantum state of the photon pairs $(S_{1},I_{1})$ and $(S_{2},I_{2})$ produced by both sources ($NL1$ and $NL2$) is a superposition of the states produced by each source which can be expressed as
\begin{align} \label{Q-State-ZWM}
|\psi\rangle &= c_{1} \ \hat{a}_{S_{1}}^{\dagger} \hat{a}_{I_{1}}^{\dagger}|vac\rangle + c_{2} \ \hat{a}_{S_{2}}^{\dagger} \hat{a}_{I_{2}}^{\dagger}|vac\rangle, \cr  
&= c_{1} |S_{1},I_{1}\rangle \ + \ c_{2} | S_{2},I_{2}\rangle, 
\end{align}
such that $|c_{1}|^{2} + |c_{2}|^{2} = 1$; $|c_{1}|^{2}$ and $|c_{2}|^{2}$ denote the emission probabilities of $NL1$ and $NL2$, respectively; $\hat{a}_{S_{j}}^{\dagger}|0\rangle_{S} = |1_{S_{j}}\rangle \equiv |S_{j}\rangle$ and $\hat{a}_{I_{j}}^{\dagger}|0\rangle_{I} = |1_{I_{j}}\rangle \equiv |I_{j}\rangle $ where $j = 1,2$.
\par
The idler $I_{1}$ from $NL1$ is aligned to the idler $I_{2}$ from $NL2$. The paths of idlers from both sources are aligned such that is not possible to ascertain the ``which-source'' information corresponding to the signal photon (that is detected). This erasure of ``which-source'' information or path indistinguishability is necessary to induce coherence in the signal photon resulting in single-photon interference. Since both crystals are weakly pumped, the possibility of stimulated emission will be negligible and furthermore, multi-photon emissions will be suppressed. The dominant process in the low-gain regime will be the two-photon emission. This phenomenon is called \textit{induced coherence without induced emission} \cite{wang1991induced}. The mathematical condition to align $I_{1}$ with $I_{2}$ can be formally written down as
\begin{align} \label{alignment condition}
\hat{a}_{I_{2}} &= e^{i \phi_{I}} \left(T \hat{a}_{I_{1}} + R \hat{a}_{0}\right),    
\end{align}
where $\phi_{I}$ is the phase picked up by $I_{1}$ while propagating from $NL1$ to $NL2$, $\hat{a}_{o}$ represents the vacuum mode at the unused port of the beamsplitter (that mathematically represents the filter) and $T$ is the transmission coefficient of the filter placed between the two sources [see Fig. (\ref{fig:ZWM-setup})] such that $|T|^{2} + |R|^{2} = 1$; $T$ is in general a complex quantity expressed as $T = |T| \ \text{exp} (i \phi_{T})$. From Eq.~(\ref{alignment condition}), it can be seen that $|T| = |\langle I_{1}|I_{2}\rangle|$, i.e., the transmission coefficient is simply the overlap between the two idler modes.  The scenario $|T| = 1 $ corresponds to perfect alignment in which case there is no ``which-source'' information available whatsoever and $|T| = 0$ corresponds to complete misalignment of idlers whereby there is total knowledge of the source of the signal photon. On substituting Eq.~(\ref{alignment condition}) in Eq.~(\ref{Q-State-ZWM}), the quantum state after enforcing the alignment condition turns to be 
\begin{align} \label{Q-State-ZWM-Aligned}
|\psi\rangle &= c_{1} \ |S_{1},I_{1}\rangle \ + \ c_{2}\ \text{e}^{-i(\phi_{I} + \phi_{T})} \ |T|\ |S_{2},I_{1}\rangle \ + \ c_{2} \ \text{e}^{-i (\phi_{I} + \phi_{R})} \ |R| \ |S_{2},I_{o}\rangle,   \end{align}
where $R = |R| \ \text{exp}(i \phi_{R})$ and $|I_{o}\rangle = \hat{a}_{o}^{\dagger}|vac\rangle$. The positive frequency part of the quantized signal field at the detector ($D$) is represented by
\begin{align} \label{Detector Operator}
 \hat{E}_{S}^{(+)} &= \hat{a}_{S_{1}} + i \ e^{i \phi_{S}} \ \hat{a}_{S_{2}},   
\end{align}
where $\phi_{S}$ is the phase difference between the two signal photons ($S_{1}$ and $S_{2}$) due to propagation. The photon counting rate ($\mathcal{R}$) at the detector ($D$) is calculated using Glauber's formula \cite{glauber1963quantum}: $\mathcal{R} \propto \langle \psi|\hat{E}_{S}^{(-)}\hat{E}_{S}^{(+)}|\psi \rangle, $ where $\hat{E}_{S}^{(-)} = \left\{\hat{E}_{S}^{(+)} \right \}^{\dagger}$. From Eqs.~(\ref{Q-State-ZWM-Aligned}) and (\ref{Detector Operator}), it follows from Glauber's formula that the single-photon counting rate (intensity) is given by
\begin{align} \label{Photon Counting Rate}
\mathcal{R} \propto  |c_{1}|^{2} + |c_{2}|^{2} + 2 |c_{1}||c_{2}| |T|\ \text{cos}(\phi_{S}+\phi_{I} + \phi_{T} + \phi'),    
\end{align}
where $\phi' = \text{arg}(c_{1}^{*}c_{2})$. The visibility ($\mathcal{V}$) of the interference pattern can be calculated from the standard formula: $\mathcal{V} = \left(\mathcal{R}_{max} - \mathcal{R}_{min} \right)/\left(\mathcal{R}_{max}+\mathcal{R}_{min} \right)$, which yields
\begin{align} \label{Visibility}
\mathcal{V} &= 2 |c_{1}||c_{2}||T|.    
\end{align}
Visibility gives a measure of the contrast in the interference fringes. It is evident from Eq.~(\ref{Visibility}) that the visibility is maximum: $\mathcal{V} = 1$, when $|c_{1}| = |c_{2}| = 1/\sqrt{2}$ and $|T| = 1$ (perfect alignment of idlers), and it vanishes altogether ($\mathcal{V} = 0$) when there is a complete misalignment ($|T| = 0$). We follow the definition in \cite{englert1996fringe} to write down the expression for predictability, 
\begin{align} \label{Predictability}
 \mathcal{P} = \left||c_{1}|^{2} - |c_{2}|^{2} \right|,  
\end{align}
which provides information about the source that produced the photon pair. Both visibility and predictability vary between $0$ and $1$, and they satisfy the well known duality relation in Eq.~(\ref{Duality-Eq}): $\mathcal{V}^{2} + \mathcal{P}^{2} \leq 1$.
\par
When $|c_{1}| = |c_{2}| = 1/\sqrt{2}$, then $\mathcal{V} = |T|$ and $\mathcal{P} = 0$ because under this condition, the emission probabilities for both crystals to produce a photon pair are equal. This means that it is not possible to determine whether the pair was created by the $NL1$ or $NL2$. Thus, the two alternatives to produce the photon pair are indistinguishable. As a result, the predictability vanishes. If the filter placed between the two crystals has a perfect transmission ($|T| = 1$), the visibility of the interference pattern takes the maximum value ($\mathcal{V} = 1$). This corresponds to the case of complete absence of ``which-way'' information. 
\par
On the other hand, when $|c_{1}| = 1$ and $|c_{2}| = 0$, or $|c_{1}| = 0$ and $|c_{2}| = 1$, then $\mathcal{V} = 0$ and $\mathcal{P} = 1$. This corresponds to the scenario where only one crystal (either $NL1$ or $NL2$) produces a photon pair which then reveals the source that emits the pair. In such a situation, the “which-source” information of the photon pair becomes completely known. As a result, interference vanishes completely, and the predictability is maximum. 
\par
In general, when $|c_{1}| \neq |c_{2}|$ (i.e., $0 < |c_{1,2}| < 1$) and $0 < |T| < 1$, both visibility and predictability will be non-zero satisfying the duality inequality in Eq.~(\ref{Duality-Eq}).
%\vspace{0.4cm}
\subsection{Mode entanglement and the persistence of triality relation}
\label{subsec: Q-ICI-Triality}

In Sec.~(\ref{subsec:Q-ICI-General}), we discussed the conceptual framework of quantum induced coherence interferometer, and determined the expressions for visibility and predictability. Here, we will take our discussion further and quantify entanglement between the signal and idler modes through the formula for I-concurrence \cite{rungta2001universal}, and explicitly demonstrate the persistence of the triality relation given by Eq.~(\ref{Triality-Eq}). In our problem, while the two subsystems are still signal and idler (that form the bi-partition), we are not discussing entanglement in the internal degrees of freedom of the two photons such as polarization, spatial variables, OAM etc. We are rather interested in the mode entanglement between the two signal path alternatives ($S_{1} $ and $S_{2}$) and the distinguishable idler modes ($I_{1}$ and $I_{2}$). The distinguishable idler modes in the induced coherence interferometer play a role analogous to the detector degrees of freedom in the abstract version of Young's double-slit setup \cite{qureshi2021predictability}.The two-photon density operator can be written down using Eq.~(\ref{Q-State-ZWM}), $\hat{\rho} = |\psi \rangle \langle \psi|$, which takes the following form:
\begin{align} \label{Density Op-ZWM}
 \hat{\rho} &= |c_{1}|^{2} |S_{1},I_{1}\rangle \langle S_{1},I_{1}| + \  |c_{2}|^{2} |S_{2},I_{2}\rangle \langle S_{2},I_{2}| \ + \ c_{1} c_{2}^{*}|S_{1},I_{1}\rangle \langle S_{2},I_{2}|\ + \ c_{1}^{*}c_{2} |S_{2},I_{2}\rangle \langle S_{1},I_{1}|.     
\end{align}
The reduced density operator for signal photons ($\hat{\rho}_{S}$) is constructed by tracing over the idler modes $I_{1}$ and $I_{2}$ in Eq.~(\ref{Density Op-ZWM}): $\hat{\rho}_{S} = \text{Tr}_{(I_{1},I_{2})}\left( \hat{\rho}\right)$, which yields
\begin{align} \label{Reduced Density Op - Signal}
 \hat{\rho}_{S} &= |c_{1}|^{2} |S_{1}\rangle \langle S_{1}| \ + \ |c_{2}|^{2} |S_{2}\rangle \langle S_{2}| \ + \ |c_{1}||c_{2}||T| e^{-i \phi} |S_{1}\rangle \langle S_{2}| \ + \  |c_{1}||c_{2}||T| e^{i \phi} |S_{2}\rangle \langle S_{1}|,   
\end{align}
where $\langle I_{1}|I_{2}\rangle = |T| \ \text{exp}(i \phi)$. In the standard $\{|S_{1}\rangle,|S_{2}\rangle\}$ basis, $\hat{\rho}_{S}$ takes the following matrix form:
\begin{align} \label{Reduced Density Op - Signal - Matrix}
 \hat{\rho}_{S} = 
 \renewcommand{\arraystretch}{1.5}
 \begin{pmatrix}
|c_{1}|^{2}  & \text{e}^{-i \phi} |c_{1} |c_{2}||T| \\
\text{e}^{i \phi} |c_{1} |c_{2}||T| & |c_{2}|^{2}
\end{pmatrix}.  
\end{align}
Using the formula for I-concurrence \cite{rungta2001universal} $\mathcal{C}_{I} = \sqrt{2 \left[1- \text{Tr}(\hat{\rho}_{s}^{2}) \right]}$, we find from Eq.~(\ref{Reduced Density Op - Signal - Matrix}) that in the induced coherence interferometer, I-concurrence is given by
\begin{align} \label{I-Concurrence - ZWM}
 \mathcal{C}_{I} &= 2 |c_{1}||c_{2}|\sqrt{1-|T|^{2}},   
\end{align}
which agrees with the result obtained using Wootters' formula for the pure state [see Eq.~(\ref{Concurrence - Wootters' formula - Result}) in Sec.~(\ref{subsec: Q-ICI-Mixed State}) and the discussion below it]. The triality relation [see Eq.~(\ref{Triality-Eq})] persists in the induced coherence interferometer. In fact, from Eqs.~(\ref{Visibility}), (\ref{Predictability}) and (\ref{I-Concurrence - ZWM}), we see that
\begin{align} \label{Triality-ZWM}
\mathcal{V}^{2} + \mathcal{P}^{2} + \mathcal{C}_{I}^{2} &= 4|c_{1}|^{2}|c_{2}|^{2}|T|^{2} \ +  \   \left||c_{1}|^{2} - |c_{2}|^{2} \right|^{2} \  + \ 4|c_{1}|^{2}|c_{2}|^{2}(1-|T|^{2}) = 1.
\end{align}
This relation is an exact quantitative trade-off among three `complementary' properties: visibility (wave-like behavior), predictability (particle-like behavior) and the degree of entanglement. When the emission probabilities of the two sources are equal: $|c_{1}|^{2} = |c_{2}|^{2}$, the predictability vanishes [see Eq.~(\ref{Predictability})] in which case, the complementarity is between visibility and I-concurrence [see Eqs.~(\ref{Visibility}) and (\ref{I-Concurrence - ZWM}), respectively]. 
\par
For a perfect idler mode overlap: $|I_{2}\rangle = |I_{1}\rangle$ ($|T| = 1$), the twin-photon state in Eq.~(\ref{Q-State-ZWM}) can be factorized into signal and idler modes viz., $|\psi\rangle = \left(c_{1} |S_{1}\rangle \ + \ c_{2} |S_{2}\rangle\right) \otimes |I_{1}\rangle $ implying that state is now separable and correspondingly $\mathcal{C}_{I} = 0$ [see Eq.~(\ref{I-Concurrence - ZWM})]. However, under the same condition, the two signal photons end up in a coherent superposition. In particular, when $|T| = 1$, the ``which-source'' information is completely erased and thus maximum amount of coherence is induced in signal photons due to perfect alignment of idlers. For the specific case when $|c_{1}| = |c_{2}| = 1/\sqrt{2}$, the visibility of the interference pattern is maximum: $\mathcal{V} = 1$. 
\par
In contrast, for a complete misalignment of idler photons, i.e., no idler mode overlap ($|T| = 0$), the twin-photon state in 
Eq.~(\ref{Q-State-ZWM}) cannot be factorized into signal and idler modes. In this situation, the state retains its form: $|\psi\rangle = c_{1} |S_{1},I_{1}\rangle \ + \ c_{2} | S_{2},I_{2}\rangle $ which clearly shows that $|\psi\rangle$ is not separable. In such a scenario I-concurrence will be non-zero. For the special case when $|c_{1}| = |c_{2}| = 1/\sqrt{2}$ and $|T| = 0$, $\mathcal{C}_{I} = 1$. I-concurrence is maximum when there is complete ``which-source'' information. The very same condition ($|T| = 0$) ensures that no coherence is induced in signal photons due to the complete availability of ``which-source'' information. As a result, in this case, the interference pattern gets washed out entirely and hence $\mathcal{V} = 0$. 

\subsection{Equivalence to the Hilbert-Schmidt coherence formalism and the triality relation from the eigenvalues of mutual coherence matrix}
\label{subsec: Q-ICI-HS Formalism}

So far, we have adopted the traditional approach to investigate the induced coherence interferometer in order to calculate visibility, predictability and I-concurrence, and verify the triality relation. The results derived in Sec.~\ref{subsec: Q-ICI-Triality} can also be obtained using the Hilbert-Schmidt coherence formalism discussed in \cite{roy2022coherence}. In this formalism, predictability, visibility and I-concurrence are computed, in the context of induced coherence interferometer, using the reduced density matrix of signal photons ($\hat{\rho}_{S}$). The formulas for the quantities in the triality relation are given by \cite{roy2022coherence}
\begin{subequations}
      \begin{align}
       \mathcal{V}^{2} &= \frac{n}{n-1} \sum_{i \neq j} |\rho_{S,ij}|^{2}, \label{Visibility-HS} \\
       \mathcal{P}^{2} &= \sum_{i=1}^{n} \rho_{S,ii}^{2} - \frac{1}{n-1}\sum_{i \neq j} \rho_{S,ii} \rho_{S,jj}, \label{Predictability-HS} \\
       \mathcal{C}_{I}^{2} &= \frac{n}{n-1} \sum_{i \neq j} \left( \rho_{S,ii} \ \rho_{S,jj}- \left| \rho_{S,ij}\right|^{2}\right), \label{I-Concurrence-HS}
      \end{align}   
     \end{subequations}
where $n=2$ for the two-source interferometric configuration that we have considered here. From Eqs.~(\ref{Visibility-HS}), (\ref{Predictability-HS}), (\ref{I-Concurrence-HS}) and (\ref{Reduced Density Op - Signal - Matrix}), we see that
{\setlength{\jot}{12pt}
\begin{subequations}
      \begin{align}
       \mathcal{V}^{2} &= 4|c_{1}|^{2}|c_{2}|^{2}|T|^{2}, \label{Visibility-HS-Result} \\
       \mathcal{P}^{2} &=  \left||c_{1}|^{2} - |c_{2}|^{2} \right|^{2}, \label{Predictability-HS-Result} \\
       \mathcal{C}_{I}^{2} &= 4|c_{1}|^{2}|c_{2}|^{2}(1-|T|^{2}), \label{I-Concurrence-HS-Result}
      \end{align}   
     \end{subequations}}
which agree with the results in Eqs.~(\ref{Visibility}), (\ref{Predictability}) and (\ref{I-Concurrence - ZWM}), respectively. From a computational perspective, it is very convenient to use the Hilbert-Schmidt coherence formalism since all the relevant quantities can be calculated, in a straightforward manner, from the reduced density matrix of the signal photon. 
\par
Interestingly, the triality relation can also be derived from the eigenvalues of mutual coherence matrix \cite{mandel_wolf_1995} for signal photons. The mutual coherence matrix ($\Gamma$), in the two-source induced coherence interferometer, is constructed the following way:
\begin{align} \label{Mutual Coherence Matrix}
 \Gamma &= 
 \renewcommand{\arraystretch}{1.5}
 \begin{pmatrix}
\langle \hat{a}_{S_{1}}^{\dagger}\hat{a}_{S_{1}}\rangle  &  \langle \hat{a}_{S_{1}}^{\dagger}\hat{a}_{S_{2}}\rangle \\
\langle \hat{a}_{S_{2}}^{\dagger}\hat{a}_{S_{1}}\rangle  & \langle \hat{a}_{S_{2}}^{\dagger}\hat{a}_{S_{2}}\rangle 
\end{pmatrix}  =
\renewcommand{\arraystretch}{1.5}
\begin{pmatrix}
|c_{1}|^{2} & \text{e}^{-i \phi} |c_{1} |c_{2}||T| \\
\text{e}^{i \phi} |c_{1} |c_{2}||T| & |c_{2}|^{2}
\end{pmatrix},
\end{align}
where the expectation values are calculated over the total quantum state in Eq.~(\ref{Q-State-ZWM}). The normalized mutual coherence (or simply the degree of coherence) between the two signal path alternatives $S_{1}$ and $S_{2}$ is defined as:
\begin{align} \label{normalized mutual coherence}
|\gamma_{12}| \equiv \frac{|\Gamma_{12}|}{\sqrt{\Gamma_{11} \Gamma_{22}}} = |T|,     
\end{align}
where $0 \leq |\gamma_{12}|\leq 1$; $|\Gamma_{12}| = |c_{1}||c_{2}||T|$, $\Gamma_{11} = |c_{1}|^{2}$ and $\Gamma_{22} = |c_{2}|^{2}$. When $|\gamma_{12}| = 0$, there is no coherence and hence no interference. The case $|\gamma_{12}| =1$ corresponds to maximum  coherence and thus the best contrast in interference fringes can be expected, in principle. The extent of coherence induced in the signal photon is determined by the alignment of idlers that is controlled by $|T|$. This directly relates mutual coherence to the absence of ``which-way'' information or path indistinguishability. We will revisit this point and elaborate it further in Sec.~(\ref{subsec: Q-ICI-Path Indistinguishability}) where we quantitatively establish the relationship between the degree of coherence and path indistinguishability, and see how these two are related to the alignment of idlers. 
\par
It can be seen that the mutual coherence matrix ($\Gamma$) in Eq.~(\ref{Mutual Coherence Matrix}) is identical to the reduced density matrix of signal photons ($\hat{\rho}_{S}$) in Eq.~(\ref{Reduced Density Op - Signal - Matrix}). This equivalence is because both $\Gamma$ and $\hat{\rho}_{S}$ provide the same information in the induced coherence interferometer. In the mutual coherence matrix ($\Gamma$), the diagonal elements correspond to intensities (or number of photons) in the paths of signal photons $S_{1}$ and $S_{2}$ and the off-diagonal elements correspond to the coherence between the two signal-path alternatives. Similarly, $\hat{\rho}_{S}$ gives all the information about population and coherence: the diagonal elements give the photon numbers corresponding to $S_{1}$ and $S_{2}$ and the off-diagonal elements correspond to the coherence in the signal field. The eigenvalues of $\Gamma$ [see Eq.~(\ref{Mutual Coherence Matrix})] are given by
\begin{align} \label{Eigenvalues - Mutual Coherence Matrix}
\lambda_{\pm} &= \frac{1}{2} \pm \frac{1}{2} \sqrt{1- 4 |c_{1}|^{2} |c_{2}|^{2} \left(1- |T|^{2} \right)}.   
\end{align}
The square of the difference in the eigenvalues of $\Gamma$: $(\Delta \lambda)^{2} \equiv (\lambda_{+} - \lambda_{-})^{2}$, can be written down as
\begin{align} \label{Eigenvalues Difference I}
(\Delta \lambda)^{2} &= 1 - 4 |c_{1}|^{2} |c_{2}|^{2} \left(1- |T|^{2} \right) = 1 -\mathcal{C}_{I}^{2},    
\end{align}
where the second term on the right hand side of Eq.~(\ref{Eigenvalues Difference I}) is the expression for I-Concurrence [see Eq.~(\ref{I-Concurrence - ZWM})]. Using the relation $|c_{1}|^{2} + |c_{2}|^{2} = 1$ and carrying out some straightforward algebraic manipulation in Eq.~(\ref{Eigenvalues Difference I}), we get
\begin{align} \label{Eignevalues Difference II}
(\Delta \lambda)^{2} &= \left(|c_{1}|^{2} + |c_{2}|^{2} \right)^{2} - 4 |c_{1}|^{2} |c_{2}|^{2} \left(1- |T|^{2} \right) \cr
&= \left(|c_{1}|^{2} - |c_{2}|^{2} \right)^{2} + 4|c_{1}|^{2}|c_{2}|^{2}|T|^{2} = \mathcal{P}^{2} + \mathcal{V}^{2},    
\end{align}
where the terms on the right hand side of the second line in Eq.~(\ref{Eignevalues Difference II}) correspond to predictability [see Eq.~(\ref{Predictability})] and visibility [see Eq.~(\ref{Visibility})], respectively. On equating the right hand side of Eqs.~ (\ref{Eigenvalues Difference I}) and (\ref{Eignevalues Difference II}), we can derive the the triality relation: $\mathcal{P}^{2} + \mathcal{V}^{2} + \mathcal{C}_{I}^{2} = 1$, from the difference in the eigenvalues of mutual coherence matrix ($\Gamma$). 

\subsection{Relationship between the degree of path indistinguishability (or ``which-source'' information) and mutual coherence, and their influence on interference and mode entanglement}
\label{subsec: Q-ICI-Path Indistinguishability}

We will now quantify the degree of path indistinguishability (or ``which-source'' information) and relate it the normalized mutual coherence [see Eq.~(\ref{normalized mutual coherence})] in the induced coherence interferometer. Towards this endeavor, we shall follow the analysis in \cite{mandel1991coherence}. We shall see how visibility and mode entanglement (through I-concurrence) depend explicitly on the degree of path indistinguishability. 
\par
Suppose that the signal photon can be produced by $NL1$ with emission probability $|c_{1}|^{2}$ or by $NL2$ with emission probability $|c_{2}|^{2}$ and furthermore if these two possibilities are intrinsically indistinguishable, then the density operator takes the following form:
\begin{align} \label{Density Op - ID}
\hat{\rho}_{ID} = |c_{1}|^{2} \ |S_{1}\rangle \langle S_{1}| \ + \ |c_{2}|^{2} \ |S_{2}\rangle \langle S_{2}| \ + \ c_{1} c_{2}^{*} \ |S_{1}\rangle \langle S_{2}| \ + \ c_{1}^{*}c_{2} \ |S_{2}\rangle \langle S_{1}|,
\end{align}
where the subscript $ID$ denotes indistinguishability. In contrast, if the  signal photon produced by either of the two sources ends up in an incoherent mixture, then the density operator takes the following diagonal form:
\begin{align} \label{Density Op - D}
\hat{\rho}_{D} = |c_{1}|^{2} \ |S_{1}\rangle \langle S_{1}| \ + \ |c_{2}|^{2} \ |S_{2}\rangle \langle S_{2}|,    
\end{align}
where the subscript $D$ denotes distinguishability. In this instance, the two possibilities to produce the signal photon are not intrinsically indistinguishable. Following the argument in \cite{mandel1991coherence}, we claim that there is a unique decomposition of $\hat{\rho}_{S}$ into $\hat{\rho}_{ID}$ and $\hat{\rho}_{D}$ by way of
\begin{align} \label{Density Op decomposition}
 \hat{\rho}_{S}  &= \mathcal{P}_{ID} \hat{\rho}_{ID} \ + \ \mathcal{P}_{D} \hat{\rho}_{D},  
\end{align}
such that $\mathcal{P}_{ID} \ + \ \mathcal{P}_{D} = 1 $; $\mathcal{P}_{ID}$ and $\mathcal{P}_{D}$ are the probabilities for the sources ($NL1$ and $NL2$) to be intrinsically indistinguishable and distinguishable, respectively. On explicitly substituting for $\hat{\rho}_{S}$, $\hat{\rho}_{ID}$ and $\hat{\rho}_{D}$ from Eqs.~(\ref{Reduced Density Op - Signal}), (\ref{Density Op - ID}) and (\ref{Density Op - D}), respectively, in Eq.~(\ref{Density Op decomposition}), and equating the corresponding matrix elements on both sides, we obtain
\begin{align} \label{Degree of Indistinguishability}
\mathcal{P}_{ID} &= |T| = |\langle I_{1}|I_{2}\rangle|.
\end{align}
From Eqs.~(\ref{normalized mutual coherence}) and (\ref{Degree of Indistinguishability}), we can clearly see that $\mathcal{P}_{ID} = |\gamma_{12}| = |T| =  |\langle I_{1}|I_{2}\rangle|$. This conveys quantitatively that in the induced coherence interferometer, the degree of coherence is the same as the degree of path indistinguishability, and experimentally they depend on the alignment of idlers that is controlled by the transmission coefficient of the filter ($|T|$) placed between the two sources [see Fig.~(\ref{fig:ZWM-setup})].  It is the alignment of idlers that erases the ``which-source'' information. In our interferometric configuration, mutual coherence measures the extent of indistinguishability of the two signal-path alternatives $S_{1}$ and $S_{2}$. Both visibility and I-concurrence in Eqs.~(\ref{Visibility}) and (\ref{I-Concurrence - ZWM}), respectively, can now be expressed in terms of $\mathcal{P}_{ID}$ which give the following expressions:
\begin{subequations}
\begin{align}
 \mathcal{V} &= 2|c_{1}||c_{2}|\mathcal{P}_{ID}, \label{Visibility-PID} \\
\mathcal{C}_{I} &= 2|c_{1}||c_{2}|\sqrt{(1-\mathcal{P}_{ID}^{2})} .\label{I-Concurrence-PID}
 \end{align}   
\end{subequations}
The dependence of visibility on the degree of coherence ($|\gamma_{12}|$) is not surprising. This is true even in classical optics. If the two optical fields are not mutually coherent, there won't be any interference [see the discussion below Eq.~(\ref{normalized mutual coherence})]. There is nothing inherently ``quantum'' about the relationship between visibility and $|\gamma_{12}|$. However, what is fundamentally ``quantum'' in our setup is the explicit dependence of visibility on the degree of path indistinguishability ($\mathcal{P}_{ID}$) [see Eq.~(\ref{Visibility-PID})] which has no classical analogue. This tells us that the extent to which the signal photon can exhibit interference is influenced by $\mathcal{P}_{ID}$. If $\mathcal{P}_{ID} = 0$, i.e., if the source of the signal photon can be identified, no coherence will be induced and hence, no interference will be observed. When $\mathcal{P}_{ID} = 1$, there is complete lack of knowledge about the source of the signal photon in which case we observe the best contrast in the interference fringes: the signal photon possesses maximum coherence. The relationship $\mathcal{P}_{ID} = |\gamma_{12}|$ is a quantitative evidence that in the interferometer that we are investigating in our work, path indistinguishability is an absolute necessity to induce coherence in signal photons. 
\par
The relationship between entanglement and  path indistinguishability ($\mathcal{P}_{ID}$), measured through I-concurrence, can be seen in Eq.~(\ref{I-Concurrence-PID}). When $\mathcal{P}_{ID} = 0$ and if $|c_{1}| = |c_{2}| = 1/\sqrt{2}$, $\mathcal{C}_{I} = 1$. This instance relates to a complete misalignment of idlers ($|T| = 0$) which implies that the idler modes $I_{1}$ and $I_{2}$ are orthogonal [see Eq.~(\ref{Degree of Indistinguishability})]. As a consequence, the idlers identify perfectly well which source produced the signal photon: maximum ``which-source'' information exists. Thus, the state $|\psi\rangle$ in Eq.~(\ref{Q-State-ZWM}) cannot be factorized into signal and idler modes which explains why I-concurrence is maximum. In this case, $|\psi\rangle = (1/\sqrt{2})(|S_{1},I_{1}\rangle + |S_{2},I_{2}\rangle)$ corresponds to a Bell state (in signal and idler modes) and is maximally entangled. Conversely, the case $\mathcal{P}_{ID} = 1$ indicates that there is a perfect alignment of idlers ($|T| = 1$) which means that the two idler modes $I_{1}$ and $I_{2}$ are indistinguishable, viz. $|I_{1}\rangle = |I_{2}\rangle$. The idlers do not possess any information on the ``signal-path'' alternative, i.e., which source ($NL1$ or $NL2$) produced the signal photon. In such a situation, the quantum state in Eq.~(\ref{Q-State-ZWM}) factorizes: $|\psi\rangle = \left(c_{1} |S_{1}\rangle \ + \ c_{2} | S_{2}\rangle\right) \otimes |I_{1}\rangle $. This implies that the signal and idler subsystems become separable, and as a result I-concurrence vanishes: $\mathcal{C}_{I} = 0$ when $\mathcal{P}_{ID} = 1$.
\par
The dependence of idler mode overlap on both visibility and I-concurrence has already been discussed in Sec.~(\ref{subsec: Q-ICI-Triality}). Here, we wanted to manifestly bring out the relationship between path indistinguishability and the idler mode overlap (alignment of idlers), and the central role that path indistinguishability plays in influencing interference and mode entanglement in the induced coherence interferometer.  

\subsection{Effect of an arbitrary background field}
\label{subsec: Q-ICI-Background Field}

We shall look into the induced coherence interferometer in the presence of an arbitrary background field (no specific form for the background will be assumed). Our analysis will be background agnostic. It could be either coherent or incoherent radiation, classical or non-classical state of radiation. In such a situation, the idler $I_{1}$ from $NL1$ is mixed with an arbitrary background field at the filter before they seed $NL2$. The total quantum state of the system will still be given by Eq.~(\ref{Q-State-ZWM}). However, the alignment condition will be slightly modified in this case,
\begin{align} \label{alignment condition- background}
\hat{a}_{I_{2}} &= e^{i \phi_{I}} \left(T \hat{a}_{I_{1}} + \sqrt{1-|T|^{2}} \ \hat{b}\right),    
\end{align}
where $\hat{b}$ is the mode corresponding to the background field that satisfies the bosonic commutation relation: $[\hat{b},\hat{b}^{\dagger}] = \text{I}$. On enforcing the alignment condition, the quantum state in Eq.~(\ref{Q-State-ZWM}) can be written as
\begin{align} \label{Q-State-Background}
 |\psi\rangle &= c_{1} \ |S_{1}\rangle \otimes |A \rangle + c_{2} \ |S_{2}\rangle \otimes |B \rangle,
 \end{align}
where $|A \rangle  \equiv |I_{1}, \xi \rangle$ and $|B \rangle \equiv |T| \text{e}^{-i (\phi_{T} \ + \ \phi_{I})}|I_{1}, \xi \rangle + \sqrt{1 - |T|^{2}} \ e^{- i \phi_{I}} \ \hat{b}^{\dagger}|0_{I_{1}}, \xi \rangle$; $|\xi \rangle$ is the ket corresponding to the background field. The density operator of the global state is given by
\begin{align} \label{Density Op - Background}
 \hat{\rho} &= |c_{1}|^{2} \ |S_{1}\rangle \langle S_{1}| \otimes |A\rangle \langle A| \ + \   |c_{2}|^{2} \ |S_{2}\rangle \langle S_{2}| \otimes |B\rangle \langle B|  \cr
 &+ c_{1}c_{2}^{*} \ |S_{1}\rangle \langle S_{2}| \otimes |A\rangle \langle B| \ + \ c_{1}^{*}c_{2} \ |S_{2} \rangle \langle S_{1}| \otimes |B\rangle \langle A|.
\end{align}
The reduced density operator for signal photons ($\hat{\rho}_{S}$) is obtained by tracing over idler and background modes [embedded in $|A\rangle$ and $|B\rangle$: see below Eq.~(\ref{Q-State-Background})] in Eq.~(\ref{Density Op - Background}): $\hat{\rho}_{S} = \text{Tr}_{A,B}(\hat{\rho})$, which then yields
\begin{align} \label{Reduced Density Op - Signal - Background}
 \hat{\rho}_{S} &= |c_{1}|^{2} \ |S_{1} \rangle \langle S_{1}| +|c_{2}|^{2} \left[1 + (1 -|T|^{2}) \bar{n} \right]|S_{2}\rangle \langle S_{2}| \cr
 &+ \ |c_{1}||c_{2}||T| e^{-i \phi} \ |S_{1}\rangle \langle S_{2}| \ + \  |c_{1}||c_{2}||T| e^{i \phi} \ |S_{2}\rangle \langle S_{1}|,
 \end{align}
where $\bar{n} = \langle \xi|\hat{b}^{\dagger} \hat{b}|\xi\rangle$ is the average number of photons in the background field, $\langle A|A \rangle = 1$, $\langle B | B \rangle = 1 +  \bar{n} (1 - |T|^{2})$ and $\langle A|B \rangle = |T| \ \text{exp}[-i (\phi_{T} \ +\ \phi_{I})]$; $\phi = \text{arg}(c_{2}) - \text{arg}(c_{1}) - \phi_{T} - \phi_{I}$. The normalized density matrix for signal photons in the standard $\{|S_{1}\rangle,|S_{2}\rangle \}$ basis takes the following form:
\begin{equation} \label{Reduced Density Op - Signal - Matrix - Background}
\hat{\rho}_{S} = \frac{1}{1 + |c_{2}|^{2} \ \bar{n} \ (1 - |T|^{2})} 
\setlength{\arraycolsep}{15pt}
\renewcommand{\arraystretch}{1.5}
\begin{pmatrix} 
|c_{1}|^{2}  & \text{e}^{-i \phi} |c_{1} |c_{2}||T| \\ 
\text{e}^{i \phi} |c_{1} |c_{2}||T| & |c_{2}|^{2} \left[1 + (1 - |T|^{2}) \bar{n} \right]
\end{pmatrix}.
\end{equation}
The expressions for visibility, predictability and I-concurrence, in the presence of an arbitrary background field, can be obtained using the formulas in Eqs.~(\ref{Visibility-HS}) - (\ref{I-Concurrence-HS}), which give us
{\setlength{\jot}{12pt}
\begin{subequations}
      \begin{align}
       \mathcal{V}^{2} &= \frac{4|c_{1}|^{2}|c_{2}|^{2}|T|^{2}}{\left[1 + |c_{2}|^{2} \ \bar{n} \ (1 - |T|^{2})\right]^{2}}, \label{Visibility-HS-Result- Background} \\
       \mathcal{P}^{2} &=  \frac{\left||c_{1}|^{2} - |c_{2}|^{2} \left[1 + (1 - |T|^{2}) \bar{n} \right] \right|^{2}}{\left[1 + |c_{2}|^{2} \ \bar{n} \ (1 - |T|^{2})\right]^{2}}, \label{Predictability-HS-Result- Background} \\
       \mathcal{C}_{I}^{2} &= \frac{4|c_{1}|^{2}|c_{2}|^{2}(1-|T|^{2})(1 + \bar{n})}{\left[1 + |c_{2}|^{2} \ \bar{n} \ (1 - |T|^{2})\right]^{2}}. \label{I-Concurrence-HS-Result - Background}
      \end{align}   
     \end{subequations}}
It is straightforward to verify from Eqs.~(\ref{Visibility-HS-Result- Background}) - (\ref{I-Concurrence-HS-Result - Background}) that the triality relation: $\mathcal{V}^{2} + \mathcal{P}^{2} + \mathcal{C}_{I}^{2} = 1$ is satisfied even in the presence of a background field. Interestingly, the specific nature of the background does not influence the results for visibility, predictability and I-concurrence. The background field enters the expressions for the quantities in the triality relation only through its mean photon number ($\bar{n}$). It does not matter whether the background is an incoherent state of radiation such as thermal photons, a highly non-classical state of light such as squeezed vacuum or a non-Gaussian state such as photon-added coherent state. It affects visibility, predictability and I-concurrence in the same way. Moreover, the background field is uncorrelated with both signal and idler photons since it is not produced by the source that generates the correlated signal-idler pair.
\par
When $|T| = 1$, the background field has no effect on $\mathcal{V}$, $\mathcal{P}$ and $\mathcal{C}_{I}$ regardless of how large $\bar{n}$ is, in which case, the results in Eqs.~(\ref{Visibility-HS-Result- Background}) - (\ref{I-Concurrence-HS-Result - Background}) reduce to Eqs.~(\ref{Visibility-HS-Result}) - (\ref{I-Concurrence-HS-Result}). This can be seen from the alignment condition in Eq.~(\ref{alignment condition- background}). For the case $|T| = 1$ which corresponds to a perfect idler mode overlap: $|I_{2}\rangle = |I_{1}\rangle $ (up to a phase factor), $\sqrt{1 - |T|^{2}} = 0$ implying that the optical mode associated with the background field vanishes completely: no background photon enters $NL2$. Thus, under this condition, the background field is decoupled altogether even if $\bar{n}$ is very large. A perfect idler mode overlap makes the filter completely opaque to the background field due to which it cannot mix with $I_{1}$ and thus, does not seed $NL2$. Its presence has no effect on the coherence induced in the signal photon. The background field can have a significant impact on visibility and concurrence only when  $|T| \ll 1$ and $\bar{n} \gg 1$. 
\par
When $\bar{n} \gg 1$, the system is no longer in the spontaneous emission regime. In such a scenario, stimulated emission dominates because $NL2$ is strongly seeded by the background field. Consequently, it will enhance the emission probability of $NL2$ in comparison to $NL1$ ($|c_{2}| \gg |c_{1}|$) which is why predictability goes up. In Eq.~(\ref{Predictability-HS-Result- Background}), for $\bar{n} \gg 1$, the second term in the numerator is much larger than the first: $|c_{2}|^{2} \left[1 + (1 - |T|^{2}) \bar{n} \right] \gg |c_{1}|^{2}$. In addition, $(1-|T|^{2})\bar{n} + 1 \simeq (1-|T|^{2})\bar{n}$. Likewise, in the denominator, $|c_{2}|^{2} \ \bar{n} \ (1 - |T|^{2}) + 1 \simeq |c_{2}|^{2} \ \bar{n} \ (1 - |T|^{2})$. Thus, under this condition,
\begin{align} \label{Predictability - Strong Background Limit}
 \mathcal{P} &\simeq \frac{|c_{2}|^{2} \ \bar{n} \ (1 - |T|^{2})}{|c_{2}|^{2} \ \bar{n} \ (1 - |T|^{2})}   = 1,
\end{align}
which physically means that $NL2$ emits a photon pair and the ``which-source'' information is completely available. Furthermore, a stronger emission from $NL2$ will make the quantum state to be more likely $|S_{2},I_{2}\rangle$ since the emission from $NL1$ will be extremely small in comparison. In this regime, the quantum state will no longer be in a superposition of $|S_{1},I_{1}\rangle$ and $|S_{2},I_{2}\rangle$, i.e., $|\psi\rangle \simeq c_{2} \ |S_{2},I_{2}\rangle$: $|\psi\rangle$ reduces to a product state in the large $\bar{n}$ limit and thus, entanglement vanishes. This can be seen from Eq.~(\ref{I-Concurrence-HS-Result - Background}):
\begin{align} \label{I-Concurrence - Strong Background Limit}
\mathcal{C}_{I} &\simeq  \ \frac{2 |c_{1}||c_{2}| \sqrt{\bar{n}}\sqrt{1 - |T|^{2}}}{|c_{2}|^{2} \ \bar{n} \ (1 - |T|^{2})} = 2 \left| \frac{c_{1}}{c_{2}}\right| \ \frac{1}{\sqrt{\bar{n}}} \ \frac{1}{\sqrt{1 - |T|^{2}}} \ \rightarrow 0,   
\end{align}
in the limit $\bar{n} \gg 1$ and $|c_{1}/c_{2}| \ll 1$. When stimulated emission dominates spontaneous emission in $NL2$ due to strong seeding from the background field, $I_{2}$ mode will be contaminated by the background photons. This will affect the mode matching between $I_{1}$ and $I_{2}$ which manifests as misalignment of idlers. As a result, lesser amount of coherence gets induced in the signal photon and hence, the visibility degrades substantially which can be seen from Eq.~(\ref{Visibility-HS-Result- Background}):
\begin{align} \label{Visibility - Strong Background Limit}
 \mathcal{V} &\simeq \frac{2 |c_{1}||c_{2}| |T|}{|c_{2}|^{2} \ \bar{n} \ (1 - |T|^{2})} = 2 \left| \frac{c_{1}}{c_{2}}\right| \ \frac{1}{\bar{n}} \ \frac{|T|}{(1- |T|^{2})} \ \rightarrow 0,  
\end{align}
in the limit $\bar{n} \gg 1$ and $|c_{1}/c_{2}| \ll 1$. Besides, when stimulated emission is strong in $NL2$, the intensity of $S_{2}$ will be much larger than $S_{1}$. When two arms of an interferometer have vastly dissimilar intensities, the visibility gets affected. This is true in both classical and quantum interferometry.   
\subsection{Mixed state in the induced coherence interferometer}
\label{subsec: Q-ICI-Mixed State}

We shall now investigate the two-photon mixed state in the induced coherence interferometer. Such a state is produced when an incoherent pump is used to illuminate the two nonlinear crystals. A two-photon mixed state can be expressed in it most general form as
\begin{align} \label{Two-photon mixed state}
\hat{\rho} &= |c_{1}|^{2}|S_{1},I_{1}\rangle \langle S_{1},I_{1}| \ + \  |c_{2}|^{2}|S_{2},I_{2}\rangle \langle S_{2},I_{2}| \ + \ \eta \ c_{1} c_{2}^{*} |S_{1},I_{1}\rangle \langle S_{2},I_{2}| \cr
&+ \ \eta \ c_{1}^{*}c_{2} |S_{2},I_{2} \rangle \langle S_{1},I_{1}|,   
\end{align}
where $|c_{1}|^{2} + |c_{2}|^{2} = 1$ and $0 \leq \eta \leq 1$. When $\eta = 1$, the density operator ($\hat{\rho}$) represents a pure state and Eq.~(\ref{Two-photon mixed state}) reduces to Eq.~(\ref{Density Op-ZWM}). On the other hand when $\eta = 0$, $\hat{\rho}$ represents a completely mixed state [no coherence in Eq.~(\ref{Two-photon mixed state})]. On enforcing the alignment condition from Eq.~(\ref{alignment condition}) in Eq.~(\ref{Two-photon mixed state}), we get
\begin{align} \label{Mixed state density op - aligned}
 \hat{\rho} &= |c_{1}|^{2} \ |S_{1},I_{1}\rangle \langle S_{1},I_{1}| \ + \ |c_{2}|^{2} ||T|^{2}|S_{2},I_{1} \rangle \langle S_{2},I_{1}| \ + \ |c_{2}|^{2}(1 - |T|^{2}) |S_{2},I_{o}\rangle \langle S_{2},I_{o}| \cr
 &+ |c_{2}|^{2}|T| \sqrt{1 - |T|^{2}} \ e^{-i \phi_{T}} |S_{2},I_{1}\rangle \langle S_{2},I_{o}| \ + \ |c_{2}|^{2}|T| \sqrt{1 - |T|^{2}} \ e^{i \phi_{T}} |S_{2},I_{o}\rangle \langle S_{2},I_{1}| \cr
 &+ \eta |c_{1}||c_{2}||T| \ e^{-i \phi'}|S_{1},I_{1}\rangle \langle S_{2},I_{1}| \ + \  \eta |c_{1}||c_{2}||T| \ e^{i \phi'}|S_{2},I_{1}\rangle \langle S_{1},I_{1}| \cr
 &+ \eta |c_{1}||c_{2}|\sqrt{1 - |T|^{2}} \ e^{-i \phi''}|S_{1},I_{1}\rangle \langle S_{2},I_{o}| \ + \ \eta |c_{1}||c_{2}|\sqrt{1 - |T|^{2}} \ e^{i \phi''}|S_{2},I_{o}\rangle \langle S_{1},I_{1}|,
\end{align}
where $\phi' = \text{arg}(c_{2}) - \text{arg}(c_{1}) - \phi_{T} - \phi_{I}$ and $\phi'' =  \text{arg}(c_{2}) - \text{arg}(c_{1}) - \phi_{I}$. In the computational basis $\{|S_{1},I_{1}\rangle, |S_{1},I_{o}\rangle, |S_{2},I_{1}\rangle, |S_{2},I_{o}\rangle \}$, $\hat{\rho}$ takes the following form:
\begin{align} \label{Mixed State Density Op - Matrix}
\hat{\rho} &=  
\setlength{\arraycolsep}{10pt}
\renewcommand{\arraystretch}{1.3}
\begin{pmatrix} 
|c_{1}|^{2}  & 0 & \text{e}^{-i \phi' }\eta |c_{1}||c_{2}||T|  & \text{e}^{-i \phi''} \eta |c_{1} |c_{2}|\sqrt{1-|T|^{2}} \\ 
0 & 0 & 0 & 0 \\
\text{e}^{i \phi' }\eta |c_{1}||c_{2}||T| & 0 & |c_{2}|^{2}|T|^{2} & \text{e}^{-i \phi_{T}}|c_{2}|^{2}|T| \sqrt{1 - |T|^{2}} \\
\text{e}^{i \phi''} \eta |c_{1} |c_{2}|\sqrt{1-|T|^{2}} & 0 &  \text{e}^{i \phi_{T}}|c_{2}|^{2}|T| \sqrt{1 - |T|^{2}} & |c_{2}|^{2}(1 - |T|^{2})
\end{pmatrix}.
\end{align}
The amount of entanglement in the two-photon mixed state [see Eq.~(\ref{Two-photon mixed state})] is quantified by concurrence using Wootters' formula \cite{wootters1998entanglement}. In order to calculate the concurrence, we need to first find out the spin-flipped state ($\tilde{\hat{\rho}}$):
\begin{align} \label{spin-flipped state}
\tilde{\hat{\rho}} &= \left(\hat{\sigma}_{y} \otimes \hat{\sigma}_{y} \right) \hat{\rho}^{*} \left(\hat{\sigma}_{y} \otimes \hat{\sigma}_{y} \right),    
\end{align}
where $\hat{\sigma}_{y}$ is the Pauli Y-operator and $\hat{\rho}^{*}$ refers to complex conjugation of each element in $\hat{\rho}$. The concurrence is then determined from the eigenvalues of $\hat{\rho} \tilde{\hat{\rho}}$. If $\lambda_{1}, \lambda_{2}, \lambda_{3}$ and $\lambda_{4}$ are the square roots of the eigenvalues of $\hat{\rho} \tilde{\hat{\rho}}$ in the decreasing order, then the concurrence of $\hat{\rho}$ is given by
\begin{align} \label{Concurrence - Wootters' formula}
 \mathcal{C}(\hat{\rho}) &= \text{max} \{\lambda_{1} - \lambda_{2} - \lambda_{3} - \lambda_{4},0 \}.  
\end{align}
From Eqs.~(\ref{Mixed State Density Op - Matrix}), (\ref{spin-flipped state}) and (\ref{Concurrence - Wootters' formula}), we find that the concurrence of the two-photon mixed state is 
\begin{align} \label{Concurrence - Wootters' formula - Result}
\mathcal{C}(\hat{\rho}) &= 2 |c_{1}||c_{2}|\eta \sqrt{1 - |T|^{2}}.    
\end{align}
When $\eta  = 1$, the expression in Eq.~(\ref{Concurrence - Wootters' formula - Result}) reduces to the result for the pure state case in Eq.~(\ref{I-Concurrence - ZWM}). However, when $\eta = 0$, $\mathcal{C}(\hat{\rho}) = 0$: the concurrence vanishes for a completely mixed state. This arises from the fact that if $\eta = 0$, the state in Eq.~(\ref{Two-photon mixed state}) becomes separable: $\hat{\rho} =  |c_{1}|^{2}|S_{1},I_{1}\rangle \langle S_{1},I_{1}| \ + \  |c_{2}|^{2}|S_{2},I_{2}\rangle \langle S_{2},I_{2}| \equiv \sum_{j} p_{j} \ \hat{\rho}_{S}^{j}  \otimes  \hat{\rho}_{I}^{j}$ (which is the mathematical condition for separability), where $p_{j} = |c_{j}|^{2}$, $\hat{\rho}_{S}^{j} = |S_{j}\rangle \langle S_{j}|$ and $\hat{\rho}_{I}^{j} = |I_{j}\rangle \langle I_{j}|$; $j = 1,2$. 
\par
The reduced density operator for signal photons $\hat{\rho}_{S, \text{mixed}}$ is constructed by tracing over the idler modes $I_{1}$ and $I_{o}$ in Eq.~(\ref{Mixed state density op - aligned}), i.e., $\hat{\rho}_{S, \text{mixed}} = \text{Tr}_{(I_{1},I_{o})}\left( \hat{\rho}\right)$:
\begin{align} \label{Mixed State Reduced Density Op - Signal}
\hat{\rho}_{S, \text{mixed}} &=   |c_{1}|^{2} |S_{1}\rangle \langle S_{1}| \ + \ |c_{2}|^{2} |S_{2}\rangle \langle S_{2}| \ + \ |c_{1}||c_{2}||T| \eta \ e^{-i \phi'} |S_{1}\rangle \langle S_{2}| \cr
&+ \  |c_{1}||c_{2}||T| \eta \ e^{i \phi'} |S_{2}\rangle \langle S_{1}|.   
\end{align}
The photon counting rate at the detector can be calculated using the standard formula: $\mathcal{R} \propto \text{Tr} \left(\hat{\rho}_{S,\text{mixed}} \hat{E}_{S}^{(-)}\hat{E}_{S}^{(+)}\right)$ where $\hat{E}_{S}^{(+)}$ is written down in Eq.~(\ref{Detector Operator}). On using Eqs.~(\ref{Mixed State Reduced Density Op - Signal}) and (\ref{Detector Operator}), it follows from this formula that
\begin{align} \label{Photon Counting Rate - Mixed State}
 \mathcal{R} \propto  |c_{1}|^{2} + |c_{2}|^{2} + 2 |T||c_{1}||c_{2}| \eta \ \text{cos}(\phi_{S} + \phi').      
\end{align}
 The visibility ($\mathcal{V}$) of the interference pattern can be determined from the standard formula: $\mathcal{V} = \left(\mathcal{R}_{max} - \mathcal{R}_{min} \right)/\left(\mathcal{R}_{max}+\mathcal{R}_{min} \right)$, which gives
\begin{align} \label{Visibility - Mixed State}
\mathcal{V} &= 2 |c_{1}||c_{2}||T| \eta.    
\end{align}
For $\eta = 1$, Eq.~(\ref{Visibility - Mixed State}) reduces to Eq.~(\ref{Visibility}) which corresponds to the pure state case and for a completely mixed state ($\eta = 0$), the visibility is zero. When the global two-photon state is completely mixed [set $\eta = 0$ in Eq.~(\ref{Two-photon mixed state})], the alignment of idlers by itself cannot induce coherence in the signal photon. While alignment can make the two signal-path alternatives indistinguishable thereby erasing the path information, it cannot create coherence in the signal photon when none exists in the global two-photon state. The process of alignment can only transfer the existing coherence in the two-photon state into the signal subsystem. This can be seen in Eq.~(\ref{Mixed State Reduced Density Op - Signal}) by setting $\eta = 0$. When the two-photon state is completely mixed, the reduced density operator of signal photons will also be diagonal (with no coherence: off -diagonal terms will be zero) due to which the interference effects vanish. The parameter $\eta$ quantifies the amount of coherence present in the global two-photon state. For $0 < \eta < 1$, the coherence is only partially preserved. Consequently, the visibility of the interference fringes will be reduced.  
\par
The predictability for the mixed state scenario will still be given by Eq.~(\ref{Predictability}) since it is related to the source that produces the photon pair. From Eqs.~(\ref{Predictability}), (\ref{Visibility - Mixed State}) and (\ref{Concurrence - Wootters' formula - Result}), the sum of the squares of visibility, predictability and concurrence turns out to be
\begin{align} \label{Triality - Mixed State}
\mathcal{V}^{2} + \mathcal{P}^{2} + \mathcal{C}^{2} \ &= \ 1 \ - \ 4|c_{1}|^{2}|c_{2}|^{2}(1 - \eta^{2}) \ \leq 1.     
\end{align}
For $\eta = 1$ (pure state), the inequality in Eq.~(\ref{Triality - Mixed State}) reduces to a tight equality given by Eq.~(\ref{Triality-ZWM}). In general, for a mixed state, $\mathcal{V}^{2} + \mathcal{P}^{2} + \mathcal{C}^{2} < 1$. The actual value of the sum in Eq.~(\ref{Triality - Mixed State}) will depend on $\eta$. For the specific case when both crystals are equally likely to emit a pair ($|c_{1}|^{2} = |c_{2}|^{2} = 1/2$), $\mathcal{V}^{2} + \mathcal{P}^{2} + \mathcal{C}^{2} = \eta^{2}$. In this situation, predictability is zero since both crystals have equal emission probabilities. The triality relation will then involve only visibility and concurrence, and the sum $\mathcal{V}^{2} + \mathcal{C}^{2}$ will once again be dictated by the value of $\eta$.

\section{Role of Alignment in Classical and Quantum Induced Coherence Interferometers}
\label{sec:Classical ZWM}

The alignment of idlers is common to both classical and quantum versions of induced coherence interferometers. Even in the classical setup, $I_{1}$ from $NL1$ seeds $NL2$. The schematic for the classical configuration is similar to Fig.~(\ref{fig:ZWM-setup}) except that in this case, the pump is strong, and ($S_{1},I_{1}$) and ($S_{2},I_{2}$) will be correlated classical beams. We shall discuss, in this section, the similarities and differences in the alignment of idlers between the two configurations viz., classical and quantum. The expressions for visibility and predictability in a classical interferometer (irrespective of the specific type of setup used) are given by \cite{saleh2025quantum}    
\begin{subequations}
\begin{align}
 \mathcal{V} &= \frac{2 \sqrt{I_{S_{1}}I_{S_{2}}}}{I_{S_{1}}+I_{S_{2}}} |\gamma_{12}|, \label{Visibility-Classical} \\
\mathcal{P} &= \frac{\left|I_{S_{1}} - I_{S_{2}}\right|}{I_{S_{1}} + I_{S_{2}}}, \label{Predictability-Classical}
 \end{align}   
\end{subequations}
where $I_{S_{1}}$ and $I_{S_{2}}$ are the intensities of signal beams in the two arms of the interferometer (in our configuration, the signal beams are detected), and $|\gamma_{12}|$ is the normalized mutual coherence. The mutual coherence, in this context, is the correlation between optical fields at two different space-time points. The predictability ($\mathcal{P}$), in a classical optical setting, is the normalized difference between the intensities of the two beams. Both visibility and predictability satisfy the well known duality relation:
\begin{align} \label{Duality-Classical}
\mathcal{V}^{2} + \mathcal{P}^{2} &= 1 \ - \ \frac{4 I_{S_{1}}I_{S_{2}}}{(I_{S_{1}}+ I_{S_{2}})^{2}} \left(1 - |\gamma_{12}|^{2} \right) \ \leq \ 1.  
\end{align}
The inequality in Eq.~(\ref{Duality-Classical}) reduces to an equality when there is perfect coherence between the two interfering fields ($|\gamma_{12}| = 1$). The correlated twin beams in the classical induced coherence interferometer are produced through the same spontaneous parametric down-conversion. In the classical theory of three-wave mixing, $E_{S} \propto E_{I}^{*}$ where $E_{S}$ and $E_{I}$ are the complex optical fields corresponding to signal and idler beams, respectively (see \cite{boyd2020nonlinear} for a pedagogical discussion on nonlinear optical processes) and the asterisk ($*$) refers to complex conjugation. It follows that $E_{S_{2}} \propto E_{I_{2}}^{*}$ where $E_{S_{2}}$ and $E_{I_{2}}$ are the fields corresponding to signal and idler, respectively, produced by $NL2$. Since, $I_{1}$ seeds $NL2$ in our setup, $E_{I_{2}} = T E_{I_{1}}$ where $T$ is the transmission coefficient of the filter placed between the two sources. Accordingly, $\langle E_{S_{1}}^{*}E_{S_{2}}\rangle \propto T \langle |E_{I_{1}}|^{2} \rangle$ which then implies that $|\gamma_{12}| \propto |T|$ since $|\gamma_{12}| \propto |\langle E_{S_{1}}^{*}E_{S_{2}}\rangle|$. This tells us that if $I_{1}$ is completely blocked by the filter ($|T| = 0$), the mutual coherence between the signal fields vanishes ($|\gamma_{12}| = 0$) as a result of which interference disappears and the visibility drops to zero [see Eq.~(\ref{Visibility-Classical})]. When $|T| = 0$, the phase coherence is not preserved between $I_{1}$ and $I_{2}$ and thus, it is not transferred to signal fields which then destroys the interference pattern. 
\par
In both classical and quantum induced coherence setups, the alignment of idlers ensures mode matching. In an experiment (regardless of whether the source is classical or quantum), alignment guarantees that both idlers have identical spatial and temporal modes, polarization etc. Furthermore, in both cases, the pump provides the stable phase reference. In the classical setting, if the same idler field drives both crystals through a perfect alignment, the signal beams remain phase locked (i.e., they will preserve the phase coherence which is essential for interference). On the other hand, if $I_{1}$ is blocked or if there is a complete misalignment, the phase coherence will no longer be preserved and thus, interference fades away. The transmission coefficient of the filter inserted between the crystals determines how good the phase locking is between the two signal beams.
\par
In the quantum induced coherence interferometer, the alignment of idlers accomplishes something more fundamental that does not have a classical analogue. It erases the “which-source” information (or creates path indistinguishability), i.e., it makes the two possible alternatives to generate signal photons indistinguishable. It is this erasure of “which-source” information that induces coherence in signal photons due to which interference is observed. A complete knowledge of ``which-source'' information destroys the coherence in the signal photon and hence no interference will be observed. There is no concept of path indistinguishability in classical optics. The quantitative relationship between path indistinguishability, mutual coherence and the transmission coefficient in the quantum setup has been discussed, in detail, in Sec.~(\ref{subsec: Q-ICI-Path Indistinguishability}). In the quantum case, the mutual coherence gives a measure of how indistinguishable the two signal path alternatives ($S_{1}$ and $S_{2}$) are, which is radically different from the notion of mutual coherence in the classical configuration.  A practical implementation of coherence induced in a classical configuration is classical imaging with undetected light \cite{cardoso2018classical}. The transmission coefficient ($|T|$) determines the extent to which path indistinguishability is accomplished and thus how much coherence is induced in the signal photon. 

\section{Conclusion}
\label{sec:concl}

In conclusion, our work reveals quantitatively how coherence is induced in signal photons by erasing the “which-source” information. Consequently, the effects of interference vanish when photon paths become identifiable. A fundamental concept that is unique to the quantum version of the induced coherence setup is the equivalence between path indistinguishability (or intrinsic indistinguishability of the sources) and the mutual coherence of signal photons. In other words, the mutual coherence in the quantum setup measures the extent of indistinguishability of signal paths. This is inherently very different from the notion of mutual coherence in classical optics which quantifies the correlation between complex optical fields at two different space-time points. Besides, we have observed that path indistinguishability plays a complementary role in affecting interference and mode entanglement. More importantly, the origin of interference and mode entanglement is the coherent superposition of the two photon-pair-generation alternatives. The quantum coherence in the global state manifests either as first-order coherence (interference) of signal photons or as entanglement between the signal and idler modes. 
\par
Our motivation, here, has been to explain the role of coherence in both classical and quantum induced coherence interferometers, and how they affect the quantities in the triality relation viz., predictability, visibility and concurrence. In addition, our investigation also elucidates the role played by the alignment of idlers in both classical and quantum models of induced coherence interferometers. While the process of alignment in both cases ensures that the two idler fields have identical spatial and temporal modes etc., and that phase coherence of the idler fields is transferred to the signal fields which is necessary to observe interference, there is something more cardinal that happens in the quantum case viz., alignment erases the “which-source” information which is unique to quantum systems. We have shown the persistence of the triality relation in the quantum induced coherence interferometer and the classical induced coherence interferometer retrieves the familiar duality inequality.
\par
All the work done so far on complementarity is in the single-mode framework. It remains to be seen how the triality relation will get modified if the multimode nature of photons is explicitly taken into account. More specifically, even within the framework of induced coherence interferometry, when the pump waist, crystal length, finite bandwidth and duration of correlated photons are considered, one will have to work out the consequences to find out the changes to the triality relation. Another area that has not been investigated so far is the complementarity of wave-particle duality for higher dimensional quantum states in the induced coherence setup. In particular, states that are produced through entanglement by path identity \cite{krenn2017entanglement, kysela2020path} are still unexplored candidates to study complementarity of wave-particle duality.  
\par
It should be pointed out that path indistinguishability is not the only mechanism to induce coherence in one of the photons in the pair. This can also be achieved by seeding one of the beams in the correlated pair, i.e., both sources are seeded by identical laser beams \cite{ou1990coherence}. In this case, coherence in the interfering photon is induced by the phase coherence of the seeding laser. This configuration has been used in \cite{yoon2021quantitative} to experimentally verify complementarity and wave-particle duality for the model discussed in \cite{qian2020quantum}.
\par
Any specific proposal on applications of induced coherence interferometers involving a trade-off between coherence, entanglement and population must be evaluated on its own merits, and the performance of this system for the desired application will have to be comprehensively examined .

%\section{Back matter}

%Back matter sections should be listed in the order Funding/Acknowledgment/Disclosures/Data Availability Statement/Supplemental Document section. An example of back matter with each of these sections included is shown below. The section titles should not follow the numbering scheme of the body of the paper. 
\vspace{0.5cm}
\begin{backmatter}
\bmsection{Funding}
Supported by the Ministry of Education, Government of India, under the Scheme for Promotion of Academic and Research Collaboration (SPARC) (No. SPARC/2025-2026/P4343, dated 20/05/2026), the Bridge Grant from the Institution of Eminence, BHU, and the I-Hub Quantum Technology Foundation, IISER Pune, under the Chanakya Doctoral Fellowship Grant (I-HUB/DF/2022-23/04)

\bmsection{Acknowledgment}
DKM acknowledges financial support from the Ministry of Education, Government of India, under the Scheme for Promotion of Academic and Research Collaboration (SPARC) (No. SPARC/2025-2026/P4343, dated 20/05/2026), the Bridge Grant from the Institution of Eminence, BHU, and the I-Hub Quantum Technology Foundation, IISER Pune, under the Chanakya Doctoral Fellowship Grant (I-HUB/DF/2022-23/04). RKP would like to acknowledge the Council of Scientific and Industrial Research (CSIR), Government of India, for the Research Associate Fellowship.

\bmsection{Disclosures}
The authors declare no conflicts of interest.

\bmsection{Data availability} 
 No data were generated or analyzed in the presented research.

%\bmsection{Supplemental document}
%A supplemental document must be called out in the back matter so that a link can be included. For example, “See Supplement 1 for supporting content.” Note that the Supplemental Document must also have a callout in the body of the paper.

\end{backmatter}

%%%%%%%%%%%%%%%%%%%%%%% References %%%%%%%%%%%%%%%%%%%%%%%%%

%%%%%%%%%% If using BibTeX:
\bibliography{sample}

@article{Bohr1928TheQP,
  title   = {The Quantum Postulate and the Recent Development of Atomic Theory},
  author  = {Niels Bohr},
  journal = {Nature},
  volume  = {121},
  number  = {3050},
  pages   = {580--590},
  year    = {1928},
  doi     = {10.1038/121580a0}
}

@book{bohr1958atomic,
  title     = {Atomic Physics and Human Knowledge},
  author    = {Bohr, Niels},
  year      = {1958},
  publisher = {John Wiley \& Sons},
  address   = {New York}
}

@article{starke2026bohr,
  title={Bohr's complementarity},
  author={Starke, Diego S and Maziero, Jonas and Basso, Marcos LW and Qureshi, Tabish},
  journal={arXiv preprint arXiv:2605.26375},
  year={2026}
}

@article{wootters1979complementarity,
  title={Complementarity in the double-slit experiment: Quantum nonseparability and a quantitative statement of Bohr's principle},
  author={Wootters, William K and Zurek, Wojciech H},
  journal={Physical Review D},
  volume={19},
  number={2},
  pages={473},
  year={1979},
  publisher={APS}
}

@article{glauber1986amplifiers,
  title={Amplifiers, Attenuators, and Schr{\"o}dinger's Cat a},
  author={Glauber, Roy J},
  journal={Annals of the New York Academy of Sciences},
  volume={480},
  number={1},
  pages={336--372},
  year={1986},
  publisher={Wiley Online Library}
}

@article{greenberger1988simultaneous,
  title={Simultaneous wave and particle knowledge in a neutron interferometer},
  author={Greenberger, Daniel M and Yasin, Allaine},
  journal={Physics Letters A},
  volume={128},
  number={8},
  pages={391--394},
  year={1988},
  publisher={Elsevier}
}

@article{mandel1991coherence,
  title={Coherence and indistinguishability},
  author={Mandel, Leonard},
  journal={Optics letters},
  volume={16},
  number={23},
  pages={1882--1883},
  year={1991},
  publisher={Optical Society of America}
}

@article{jaeger1993complementarity,
  title={Complementarity of one-particle and two-particle interference},
  author={Jaeger, Gregg and Horne, Michael A and Shimony, Abner},
  journal={Physical Review A},
  volume={48},
  number={2},
  pages={1023},
  year={1993},
  publisher={APS}
}

@article{jaeger1995two,
  title={Two interferometric complementarities},
  author={Jaeger, Gregg and Shimony, Abner and Vaidman, Lev},
  journal={Physical Review A},
  volume={51},
  number={1},
  pages={54},
  year={1995},
  publisher={APS}
}

@article{englert1996fringe,
  title={Fringe visibility and which-way information: An inequality},
  author={Englert, Berthold-Georg},
  journal={Physical review letters},
  volume={77},
  number={11},
  pages={2154},
  year={1996},
  publisher={APS}
}

@article{qian2018entanglement,
  title={Entanglement limits duality and vice versa},
  author={Qian, X-F and Vamivakas, AN and Eberly, JH},
  journal={Optica},
  volume={5},
  number={8},
  pages={942--947},
  year={2018},
  publisher={Optical Society of America}
}

@article{qian2020turning,
  title={Turning off quantum duality},
  author={Qian, X-F and Konthasinghe, K and Manikandan, SK and Spiecker, D and Vamivakas, AN and Eberly, JH},
  journal={Physical Review Research},
  volume={2},
  number={1},
  pages={012016},
  year={2020},
  publisher={APS}
}

@article{roy2022coherence,
  title={Coherence, path predictability, and I concurrence: A triality},
  author={Roy, Abhinash Kumar and Pathania, Neha and Chandra, Nitish Kumar and Panigrahi, Prasanta K and Qureshi, Tabish},
  journal={Physical Review A},
  volume={105},
  number={3},
  pages={032209},
  year={2022},
  publisher={APS}
}

@article{qian2020quantum,
  title={Quantum duality: A source point of view},
  author={Qian, X-F and Agarwal, GS},
  journal={Physical Review Research},
  volume={2},
  number={1},
  pages={012031},
  year={2020},
  publisher={APS}
}

@article{swain2026persistence,
  title={Persistence of quantum triality relations in open n-dimensional systems},
  author={Swain, Pratidhwani and Sarkar, Ramita and Tripathy, Sukanta K and Panigrahi, Prasanta K},
  journal={APS Open Science},
  volume={1},
  pages={000006},
  year={2026},
  publisher={APS}
}

@article{benerjee2026wave,
  title={Wave-particle duality and entanglement in neutrino oscillation},
  author={Benerjee, Rajrupa and Swain, Pratidhwani and Panigrahi, Prasanta K and Patra, Sudhanwa},
  journal={Nuclear Physics B},
  pages={117509},
  year={2026},
  publisher={Elsevier}
}

@article{zou1991induced,
  title={Induced coherence and indistinguishability in optical interference},
  author={Zou, Xing-Yu and Wang, Lei J and Mandel, Leonard},
  journal={Physical review letters},
  volume={67},
  number={3},
  pages={318},
  year={1991},
  publisher={APS}
}

@misc{feynman1966feynman,
  title={The feynman lectures on physics, vol. 3: Quantum mechanics},
  author={Feynman, Richard P and Leighton, Robert B and Sands, Matthew and Lindsay, R Bruce},
  year={1966},
  publisher={American Institute of Physics}
}

@article{hochrainer2022quantum,
  title={Quantum indistinguishability by path identity and with undetected photons},
  author={Hochrainer, Armin and Lahiri, Mayukh and Erhard, Manuel and Krenn, Mario and Zeilinger, Anton},
  journal={Reviews of Modern Physics},
  volume={94},
  number={2},
  pages={025007},
  year={2022},
  publisher={APS}
}

@article{wang1991induced,
  title={Induced coherence without induced emission},
  author={Wang, Lipo J and Zou, Xy Y and Mandel, Leonard},
  journal={Physical Review A},
  volume={44},
  number={7},
  pages={4614},
  year={1991},
  publisher={APS}
}

@incollection{lahiri2020partially,
  title={Partially polarized light and complementarity in quantum mechanics},
  author={Lahiri, Mayukh},
  booktitle={Progress in Optics},
  volume={65},
  pages={313--346},
  year={2020},
  publisher={Elsevier}
}

@article{ou1990coherence,
  title={Coherence in two-photon down-conversion induced by a laser},
  author={Ou, ZY and Wang, LJ and Zou, XY and Mandel, L},
  journal={Physical Review A},
  volume={41},
  number={3},
  pages={1597},
  year={1990},
  publisher={APS}
}

@article{glauber1963quantum,
  title={The quantum theory of optical coherence},
  author={Glauber, Roy J},
  journal={Physical Review},
  volume={130},
  number={6},
  pages={2529},
  year={1963},
  publisher={APS}
}

@article{rungta2001universal,
  title={Universal state inversion and concurrence in arbitrary dimensions},
  author={Rungta, Pranaw and Bu{\v{z}}ek, Vladimir and Caves, Carlton M and Hillery, Mark and Milburn, Gerard J},
  journal={Physical Review A},
  volume={64},
  number={4},
  pages={042315},
  year={2001},
  publisher={APS}
}

@book{mandel_wolf_1995,
  author    = {Mandel, Leonard and Wolf, Emil},
  title     = {Optical Coherence and Quantum Optics},
  publisher = {Cambridge University Press},
  address   = {Cambridge},
  year      = {1995},
  doi       = {10.1017/CBO9781139644105},
  isbn      = {9780521417112}
}

@article{qureshi2021predictability,
  title={Predictability, distinguishability, and entanglement},
  author={Qureshi, Tabish},
  journal={Optics Letters},
  volume={46},
  number={3},
  pages={492--495},
  year={2021},
  publisher={Optical Society of America}
}

@article{wootters1998entanglement,
  title={Entanglement of formation of an arbitrary state of two qubits},
  author={Wootters, William K},
  journal={Physical Review Letters},
  volume={80},
  number={10},
  pages={2245},
  year={1998},
  publisher={APS}
}

@book{saleh2025quantum,
  title={Quantum Photonics},
  author={Saleh, Bahaa EA},
  year={2025},
  publisher={Springer}
}

@article{cardoso2018classical,
  title={Classical imaging with undetected light},
  author={Cardoso, Arthur Castro and Berruezo, LP and {\'A}vila, DF and Lemos, GB and Pimenta, WM and Monken, CH and Saldanha, PL and P{\'a}dua, S},
  journal={Physical Review A},
  volume={97},
  number={3},
  pages={033827},
  year={2018},
  publisher={APS}
}

@book{boyd2020nonlinear,
  title     = {Nonlinear Optics},
  author    = {Boyd, Robert W.},
  edition   = {4th},
  year      = {2020},
  publisher = {Academic Press},
  address   = {Elsevier},
  isbn      = {978-0-12-811002-7}
}

@article{yoon2021quantitative,
  title={Quantitative complementarity of wave-particle duality},
  author={Yoon, Tai Hyun and Cho, Minhaeng},
  journal={Science Advances},
  volume={7},
  number={34},
  pages={eabi9268},
  year={2021},
  publisher={American Association for the Advancement of Science}
}

@article{Liu2012_WaveParticle,
  title = {Relation between wave-particle duality and quantum uncertainty},
  author = {Liu, Hong-Yu and Huang, Jie-Hui and Gao, Jiang-Rui and Zubairy, M. Suhail and Zhu, Shi-Yao},
  journal = {Phys. Rev. A},
  volume = {85},
  issue = {2},
  pages = {022106},
  numpages = {5},
  year = {2012},
  month = {Feb},
  publisher = {American Physical Society},
  doi = {10.1103/PhysRevA.85.022106},
  url = {https://link.aps.org/doi/10.1103/PhysRevA.85.022106}
}

@article{jakob2010quantitative,
  title={Quantitative complementarity relations in bipartite systems: Entanglement as a physical reality},
  author={Jakob, Matthias and Bergou, J{\'a}nos A},
  journal={Optics Communications},
  volume={283},
  number={5},
  pages={827--830},
  year={2010},
  publisher={Elsevier}
}

@article{Jakob2007Complementarity,
  author    = {Jakob, Matthias and Bergou, Janos A.},
  title     = {Complementarity and entanglement in bipartite qudit systems},
  journal   = {Physical Review A},
  volume    = {76},
  number    = {5},
  pages     = {052107},
  year      = {2007},
  publisher = {American Physical Society},
  doi       = {10.1103/PhysRevA.76.052107},
  url       = {https://link.aps.org/doi/10.1103/PhysRevA.76.052107}
}

@article{machado2024complementarity,
  title={Complementarity relationship between coherence and path distinguishability in an interferometer based on induced coherence},
  author={Machado, Gerard J and Sendra, Lluc and Vall{\'e}s, Adam and Torres, Juan P},
  journal={Physical Review A},
  volume={110},
  number={1},
  pages={012421},
  year={2024},
  publisher={APS}
}

@article{krenn2017entanglement,
  title={Entanglement by path identity},
  author={Krenn, Mario and Hochrainer, Armin and Lahiri, Mayukh and Zeilinger, Anton},
  journal={Physical review letters},
  volume={118},
  number={8},
  pages={080401},
  year={2017},
  publisher={APS}
}

@article{kysela2020path,
  title={Path identity as a source of high-dimensional entanglement},
  author={Kysela, Jaroslav and Erhard, Manuel and Hochrainer, Armin and Krenn, Mario and Zeilinger, Anton},
  journal={Proceedings of the National Academy of Sciences},
  volume={117},
  number={42},
  pages={26118--26122},
  year={2020},
  publisher={National Academy of Sciences}
}

%%%%%%%%%% If preparing manually:

\end{document}